\documentclass[aps,10pt,pra,twocolumn,showpacs,floatfix,superscriptaddress,nofootinbib]{revtex4-2}

\usepackage{graphicx}
\usepackage{natbib}
\usepackage{dcolumn}
\usepackage{amsmath}
\usepackage{epsfig}
\usepackage{bm}
\usepackage{amssymb}
\usepackage{hyperref}
\usepackage{times}
\hypersetup{colorlinks=true,linkcolor=blue,citecolor=blue,urlcolor=blue}

\begin{document}

\title{Experimental optimal discrimination of \texorpdfstring{$N$}{N} states of a qubit with fixed rates of inconclusive outcomes}

\author{L. F. Melo}
\email{felipe-melo@ufmg.br}
\affiliation{Departamento de F\'isica, Universidade Federal de Minas Gerais, Belo Horizonte, MG, Brazil}

\author{M. A. Sol\'is-Prosser}
\affiliation{Departamento de Ciencias F\'isicas, Universidad de La Frontera, Temuco,Chile}

\author{O. Jim\'enez}
\affiliation{Centro de \'Optica e Informaci\'on Cu\'antica, Facultad de Ciencias, Universidad Mayor, Camino La Pir\'amide 5750, Huechuraba, Santiago, Chile}

\author{A. Delgado}
\affiliation{Departamento de F\'isica, Universidad de Concepci\'on, Casilla 160-C, Concepci\'on, Chile}
\affiliation{Millennium Institute for Research in Optics, Universidad de Concepci\'on, 160-C Concepci\'on, Chile}

\author{L. Neves}
\affiliation{Departamento de F\'isica, Universidade Federal de Minas Gerais, Belo Horizonte, MG, Brazil}

\date{\today}

\begin{abstract}  
In a general optimized measurement scheme for discriminating between nonorthogonal quantum states, the error rate is minimized under the constraint of a fixed rate of inconclusive outcomes (FRIO). This so-called optimal FRIO measurement encompasses the standard and well known minimum-error and optimal unambiguous (or maximum-confidence) discrimination strategies as particular cases. Here, we experimentally demonstrate the optimal FRIO discrimination between $N=2,3,5,$ and $7$ equally likely symmetric states of a qubit  encoded in photonic path modes. Our implementation consists of applying a probabilistic quantum map which  increases the distinguishability between the inputs in a controlled way, followed by a minimum-error measurement on the successfully transformed outputs. The results obtained corroborate this two-step approach and, in our experimental scheme, it can be straightforwardly extended to higher dimensions. The optimized measurement demonstrated here will be useful for quantum communication scenarios where the error rate and the inconclusive rate must be kept below the levels provided by the respective standard strategies. 
\end{abstract}

\maketitle

\section{Introduction}
The problem of quantum state discrimination has been introduced in the late 1960s and consists of designing a measurement strategy to optimally determine in which state a quantum system was prepared, given a set $\{\hat{\rho}_j\}$ of possible states with associated {\it a priori} probabilities $\{\eta_j\}$ \cite{Heltrom67-I,Heltrom67-II,Helstrom69}. If formulated in terms of a sender that transmits a message built from the ``alphabet'' $\{\hat{\rho}_j\}$ to a receiver that extracts it through  measurements, we see that this problem is the essence of quantum communication \cite{Gisin02,BarnettBook}. Furthermore, as it encompasses the measurement process in quantum theory, it naturally underlies many applications in quantum information processing and quantum foundations \cite{Chefles00,Bae15}.

When the quantum states are not mutually orthogonal, quantum theory forbids perfect discrimination between them. In this case, any discrimination strategy will have a nonzero probability of erroneous or inconclusive results \cite{Chefles00,Bae15,Barnett09,Bergou10}. In the pioneering minimum-error (ME) measurement, each outcome is used to infer the received state and the overall error probability is minimized \cite{HelstromBook,Holevo73,Yuen75}. On the other hand, the optimal unambiguous discrimination (UD) strategy enables error-free identifications of linearly independent states, with inconclusive results in a minimum fraction of trials \cite{Ivanovic87,Dieks88,Peres88,Chefles98-1}. More recently, a strategy analogous to UD was conceived to discriminate linearly dependent states. In the optimal maximum-confidence (MC) measurement \cite{Croke06}, one maximizes the probability $P(\hat{\rho}_j|j)$, taken as our \emph{confidence} in associating outcome $j$ to state $\hat{\rho}_j$, and minimizes the rate of inconclusive results. 

These state discrimination strategies were shown to be extreme and particular cases of a more general optimized measurement scheme in which the error rate is minimized under the constraint of a {\it fixed rate of inconclusive outcomes} (FRIO). The optimal FRIO strategy was proposed in Refs.~\cite{Chefles98-3,Zhang99}, and shown to be a scheme that interpolates between ME and optimal UD measurements. Thereafter, recent works also show that it interpolates between ME and optimal MC measurements when the maximum confidences are the same for all states in the set \cite{Bagan12,Herzog12}.

In addition to generalizing fundamental discrimination strategies, the optimal FRIO measurement is useful in practical situations. For instance, in a quantum communication scenario where the error rate and the inconclusive rate must be kept below the levels provided by the ME and optimal UD/MC strategies, respectively, there will be a family of optimal FRIO measurements that accomplishes those requirements. Examples of the application of FRIO in this context were given for protocols like quantum teleportation \cite{Prosser16} and dense coding \cite{Kogler17} with nonmaximally entangled states.

Considering two-dimensional spaces, all extreme strategies have been demonstrated experimentally. Implementations of the ME measurement for sets of two \cite{Barnett97}, three, and four \cite{Clarke00} states, optimal UD for sets of two states \cite{Huttner96,Clarke01}, and optimal MC for sets of three states \cite{Mosley06}, were provided. There are also demonstrations in high-dimensional discrete spaces for ME \cite{Prosser17} and optimal UD \cite{Mohseni04,Prosser21,Agnew14} measurements. Only recently the optimal FRIO strategy was implemented for discriminating two nonorthogonal pure states of a polarization-encoded qubit prepared with arbitrary {\it a priori} probabilities \cite{Gomez22}. 

In this work, we also address the optimal FRIO measurement for a qubit, but now designed to experimentally discriminate between $N$ states prepared with equal {\it a priori} probabilities. The strategy is demonstrated for $N=2,3,5,$ and $7$ states of a qubit encoded in photonic path modes. Our implementation is divided into two steps: first, using a programmable spatial light modulator, we carry out an optimal quantum state separation, a probabilistic quantum map which  increases the distinguishability between the inputs in a controlled way \cite{Chefles98}; then, the successfully transformed output states are discriminated with the ME measurement devised in \cite{Prosser17}. This approach is corroborated by the experimental results obtained, where the minimum error rates are achieved for fixed rates of inconclusive outcomes, encompassing the extreme strategies as particular cases. The experimental scheme presented here can also be straightforwardly extended for the optimal FRIO discrimination of nonorthogonal qudit states, thus consisting of an useful platform for further research in this topic and potential applications in related quantum information protocols in two- or higher dimensional spaces.


\section{An overview of state discrimination and its optimal strategies}
\label{sec:QSD}

Consider a quantum system randomly prepared in one of $N$ states $\{\hat{\rho}_j\}_{j=0}^{N-1}$ with {\it a priori} probabilities $\{\eta_j\}$ ($\sum_j\eta_j=1$). Suppose we are given this system and asked what its quantum state is. In order to answer the question, we implement a measurement on the system and use its outcome as a guide. In general, there are two classes of  outcomes: conclusive and inconclusive. The former allows us to identify the state, and this identification may be correct or not. The latter does not allow us to identify any state. This scenario can be properly described by an ($N+1$)-outcome POVM $\{\hat{\Pi}_0,\ldots\hat{\Pi}_{N-1},\hat{\Pi}_?\}$ (with $\sum_j\hat{\Pi}_j+\hat{\Pi}_?=\hat{I}$), where each element $\hat{\Pi}_{j}$ is associated with a conclusive identification of the state as $\hat{\rho}_j$, while $\hat{\Pi}_?$ is associated with an inconclusive answer. The process is characterized by the average probabilities of erroneous ($P_e$), correct  ($P_c$) and inconclusive ($Q$) results, which are given by
\begin{subequations}     \label{eq:PePcQ}
\begin{align} 
P_e&=\sum_{\substack{j,k=0\\j\neq k}}^{N-1}\eta_j{\rm Tr}(\hat{\rho}_j\hat{\Pi}_k), \\
P_c&=\sum_{j=0}^{N-1}\eta_j{\rm Tr}(\hat{\rho}_j\hat{\Pi}_j), \\
Q&=\sum_{j=0}^{N-1}\eta_j{\rm Tr}(\hat{\rho}_j\hat{\Pi}_?), 
\end{align}
\end{subequations}
and satisfy $P_e+P_c+Q=1$. Clearly, $P_e$ and $P_c$ are both related to the conclusive events.  

The goal now is to find the POVM that optimizes these probabilities according to some pre-established criterion, which will define a different measurement strategy. For instance, in the optimal FRIO measurement, the error probability $P_e$ must be minimized under the constraint that the rate of inconclusive results $Q$ has a fixed value in the range $0\leqslant Q\leqslant Q_{\rm cr}$, yielding $P^{\rm min}_e(Q)$. Here, $Q_{\rm cr}$ denotes a critical value of $Q$, above which the minimum relative error rate\footnote{This is the minimum error probability conditioned on obtaining a conclusive result for a fixed value of $Q$.} $P_e^{\rm min}(Q)/(1-Q)$ becomes a constant \cite{Bagan12,Herzog12}. 

The ME and optimal UD strategies are extreme and particular cases of the optimal FRIO measurement. In ME, where inconclusive results are not allowed, $P_e$ is minimized subject to $Q=0$ (and hence $\hat{\Pi}_?=\hat{0}$). In optimal UD, the measurement must provide the minimum rate of inconclusive outcomes $Q=Q_{\rm cr}\equiv Q^{\rm\textsc{ud}}$ subject to $P_e^{\rm min}(Q^{\rm\textsc{ud}})=0$.\footnote{Note that it is also possible to have a suboptimal UD measurement where $P_e^{\rm min}(Q)=0$ for $Q^{\rm\textsc{ud}}< Q< 1$.} The optimal MC measurement is also an extreme case of the optimal FRIO for $Q=Q_{\rm cr}\equiv Q^{\rm\textsc{mc}}$, but only when the maximum confidence $C_j\equiv \max_{\hat{\Pi}_j}[P(\hat{\rho}_j|j)]$ is the same for each of the $N$ states \cite{Herzog12}, as will be the case in this work. When these states are linearly independent, we will have $C_j=1$ $\forall j$, and the optimal MC measurement coincides with the optimal UD measurement. 

Clearly, the optimal FRIO strategy interpolates between the two extremes ME and optimal UD (or MC). Its figure of merit, $P^{\rm min}_e(Q)$, is a nonincreasing convex function \cite{Bagan12}, so that the error rate cannot increase with $Q$. Intuitively, we can expect that FRIO will reduce the optimal error rate in comparison with ME by allowing a nonzero rate of inconclusive outcomes. Similarly, it will reduce this rate in comparison with the optimal UD (or MC) measurement, at the expense of a higher error rate in the discrimination. 

The optimal strategies outlined above provide useful information with probability $1-Q$. In this sense, ME is a deterministic strategy, whereas those with $Q>0$ are probabilistic. Yet there is an intrinsic connection between them: the probabilistic strategies can be decomposed into a discrimination between conclusive and inconclusive events, followed by a ME measurement for the conclusive ones \cite{Chefles98,Jimenez07,Jimenez11,Nakahira12}. We shall explore this point in more detail in both the theoretical description and experimental implementation of the optimal FRIO discrimination in the next sections.

\section{Optimal FRIO discrimination of \texorpdfstring{$N$}{N} symmetric states of a qubit}

Deriving analytical solutions of the optimization problem posed by state discrimination is, in general, a difficult task. In particular, a set of $N$ symmetric $d$-dimensional pure states prepared with equal {\it a priori} probabilities (i.e., $\eta_j=1/N$ $\forall j$) belongs to a class of analytically solvable cases. For these states, that will be defined bellow for a qubit ($d=2$), the optimal measurement is known for ME \cite{Ban97}, optimal UD \cite{Chefles98-2}, MC \cite{Jimenez11} and FRIO \cite{Bagan12,Herzog12} strategies, which have important implications in quantum communications (e.g., see \cite{Bennett92,Phoenix00,Renes04,Neves12,Prosser14,Kogler17}). In this section, we describe the optimal FRIO measurement and obtain $P_e^{\rm min}(Q)$ for $N$ equally likely symmetric states of a qubit, focusing on its physical implementation as the two-step process mentioned above and that will be adopted in our experiment.

\subsection{Symmetric pure states of a qubit}
\begin{figure}[t]
\centerline{\includegraphics[width=1\columnwidth]{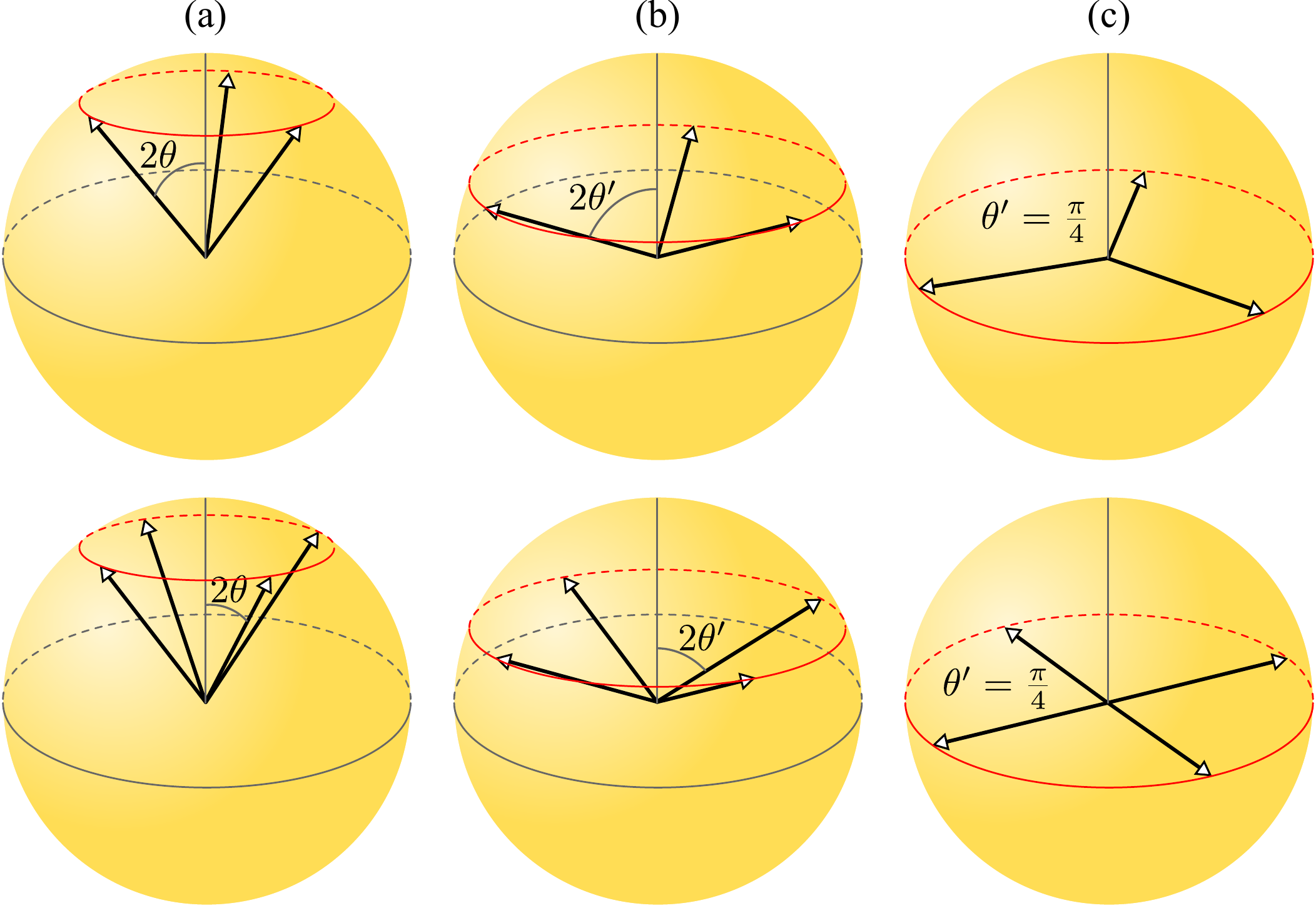}}
\caption{\label{fig:Bloch} Symmetric states of a qubit on the Bloch sphere for  $N=3$ (first row) and $N=4$ (second row). (a) Input states [Eq.~(\ref{eq:symmetric_alpha})]; (b) output states separated by $\theta<\theta'<\pi/4$ [Eq.~(\ref{eq:symmetric_beta})]; (c)  output states maximally separated [Eq.~(\ref{eq:symmetric_uniform})]. To simplify this illustration, we assumed $\varphi=0$ in Eq.~(\ref{eq:symmetric_alpha}).}
\end{figure}

Let a pure state of a qubit be written as
\begin{equation}    \label{eq:fiducial}
|\alpha_0(\theta)\rangle=\cos\theta|0\rangle+e^{i\varphi}\sin\theta|1\rangle,
\end{equation}
where $0\leqslant\theta\leqslant\pi/4$, $0\leqslant\varphi<2\pi$ and $\{|0\rangle,|1\rangle\}$ is the computational basis for its Hilbert space $\mathcal{H}$. Now, consider a unitary operation, acting on $\mathcal{H}$, given by 
$\hat{V}=|0\rangle\langle 0|+\omega|1\rangle\langle 1|$,
where $\omega=\exp(2\pi i/N)$, for some integer $N\geqslant 2$. Taking (\ref{eq:fiducial}) as a fiducial state, we can generate a set of $N$ states $\{|\alpha_j(\theta)\rangle\}_{j=0}^{N-1}$ by applying the above unitary as follows
\begin{align}
|\alpha_j(\theta)\rangle & =  \hat{V}^j|\alpha_0(\theta)\rangle \nonumber \\
&= \cos\theta|0\rangle+e^{i\varphi}\omega^j\sin\theta|1\rangle.
\label{eq:symmetric_alpha}
\end{align}
Under the action of $\hat{V}^j$ the fiducial state is rotated around the $z$-axis of the Bloch sphere by an azimuthal angle $2\pi j/N$ while keeping its polar angle $2\theta$. The $N$ states generated in this way are symmetrically distributed on the parallel of latitude $\pi/2-2\theta$ north of the Bloch sphere equator, as sketched in Fig.~\hyperref[fig:Bloch]{\ref{fig:Bloch}(a)} for $N=3$ (first row) and $N=4$ (second row). They are called \emph{symmetric states} with respect to $\hat{V}$ \cite{Ban97,Chefles98-2}; note that $\hat{V}^N=\hat{I}$, where $\hat{I}$ is the identity on the qubit space.

\subsection{Optimal separation of symmetric qubit states}
\label{subsec:QSS}

Given an input set of symmetric states $\{|\alpha_j(\theta)\rangle\}_{j=0}^{N-1}$ as defined in Eq.~(\ref{eq:symmetric_alpha}), consider, for all $j$, the transformation $|\alpha_j(\theta)\rangle\rightarrow|\beta_j(\theta')\rangle$, where
\begin{equation}    \label{eq:symmetric_beta}
|\beta_j(\theta')\rangle = \cos\theta'|0\rangle+\omega^j\sin\theta'|1\rangle,
\end{equation}
and $\theta\leqslant\theta'\leqslant\pi/4$. This transformation removes the phase\footnote{Removing this phase is not a requirement; the desired effect of the transformation would be the same whether we kept it or not. However, we choose to remove it at this point so that the theoretical description presented in this section fits our experimental implementation described in Sec.~\ref{sec:Experiment}.} $e^{i\varphi}$ and, more importantly, increases the polar angle of the inputs, generating a new set of symmetric states on a parallel of the Bloch sphere closer to the equator than before, as can be seen in Fig.~\hyperref[fig:Bloch]{\ref{fig:Bloch}(b)}. As a consequence, $|\langle\beta_i(\theta')|\beta_j(\theta')\rangle|\leqslant|\langle\alpha_i(\theta)|\alpha_j(\theta)\rangle|$ $\forall$ $i\neq j$, i.e., the transformation reduces the overlaps, making the output states more distinguishable than the input ones. For this reason, it was named \emph{state separation} \cite{Chefles98}. In particular, when $\theta'=\pi/4$, the separation is maximal and the output states become uniform: 
\begin{equation}     \label{eq:symmetric_uniform}
|\beta_j(\pi/4)\rangle\equiv|u_j\rangle = \frac{1}{\sqrt{2}}(|0\rangle+\omega^j|1\rangle).
\end{equation}
These states are located on the equator of Bloch sphere, as shown in Fig.~\hyperref[fig:Bloch]{\ref{fig:Bloch}(c)}, and they are the maximally distinguishable symmetric states (e.g., for $N=2$, they are orthogonal).

State separation is clearly a probabilistic transformation, otherwise one could apply it to render nonorthogonal states into perfectly distinguishable orthogonal ones in a deterministic way, thus contradicting the rules of quantum theory. It is characterized by two possible outcomes, success or failure: the former leads to the desired separation whereas the latter leads to output states less distinguishable than the inputs. These outcomes occur with probabilities $p_s$ and $p_f$, respectively, and the optimal transformation is the one that maximizes $p_s$ (or equivalently, minimizes $p_f$) for a prescribed separation. For $N$ equally likely symmetric states of arbitrary dimension, the maximal success probability was found in \cite{Prosser16}; in the case of a qubit, it is given by 
\begin{equation}     \label{eq:p_succ}
p_s(\theta')=\left(\frac{\sin\theta}{\sin\theta'}\right)^2=1-p_f(\theta').
\end{equation}

The corresponding measurement operators associated with these optimal probabilities were also derived in \cite{Prosser16}. Here, we describe their physical implementation which complies with our experimental realization. For this, consider an auxiliary qubit in the pure state $|{\mathsf{v}}\rangle$, which is an element of the computational basis $\{|{\mathsf{h}}\rangle=(1,0)^{\mathsf{T}},|{\mathsf{v}}\rangle=(0,1)^{\mathsf{T}}\}$ for its Hilbert space $\mathcal{H}_a$. The ancilla is attached to the main qubit in a given state $|\alpha_j(\theta)\rangle$ via tensor product, so that the input two-qubit state becomes $|\alpha_j(\theta)\rangle\otimes|{\mathsf{v}}\rangle$. Now, consider the following unitary operation acting on $\mathcal{H}\otimes\mathcal{H}_a$:
\begin{equation}    \label{eq:Unitary_coupling}
\hat{U}(\theta')=e^{i\varphi}|0\rangle\langle 0|\otimes\left[
\begin{array}{cc}
\xi(\theta') & \tau(\theta') \\[1mm]
-\tau(\theta') & \xi(\theta')
\end{array}
\right]_a+|1\rangle\langle 1|\otimes\hat{I}_a,
\end{equation}
where $\xi(\theta')=\tan\theta\cot\theta'=\sqrt{1-\tau^2(\theta')}$ and $\hat{I}_a$ is the identity on the ancilla space. By applying this operation on the input state and using Eqs.~(\ref{eq:symmetric_alpha}), (\ref{eq:symmetric_beta}) and (\ref{eq:p_succ}), it is straightforward to show that
\begin{equation}    \label{eq:Unitary}
\hat{U}(\theta')|\alpha_j(\theta)\rangle|{\mathsf{v}}\rangle = \sqrt{p_s(\theta')}|\beta_j(\theta')\rangle|{\mathsf{v}}\rangle+\sqrt{p_f(\theta')}|0\rangle|{\mathsf{h}}\rangle.
\end{equation}
After the unitary system-ancilla coupling, the protocol is concluded by a projective measurement on the ancilla's computational basis: it is projected either onto $|{\mathsf{v}}\rangle$ with the maximal success probability, thus accomplishing the state separation, or onto $|{\mathsf{h}}\rangle$, rendering the failure outputs completely indistinguishable.

\subsection{Discriminating the separated states with minimum error}

The qubit states generated by successful transformations, namely  $\{|\beta_j(\theta')\rangle\}_{j=0}^{N-1}$ given by Eq.~(\ref{eq:symmetric_beta}), also form a set of $N$ equally likely symmetric states. Assume that one wants to discriminate them with a ME strategy. The optimized measurement for this task is an $N$-outcome POVM $\{\hat{\Pi}_k^{\rm\textsc{me}}\}_{k=0}^{N-1}$ given by \cite{Ban97,Jimenez11}
\begin{equation}    \label{eq:POVM_ME}
\hat{\Pi}_k^{\rm\textsc{me}}=\frac{2}{N}|u_k\rangle\langle u_k|,
\end{equation}
where $\{|u_k\rangle\}$ are the uniform symmetric states given by Eq.~(\ref{eq:symmetric_uniform}). Denoting the average probability of error for discriminating the separated states as $p^\beta_e(\theta')$, its minimum value will be
\begin{align}
p^\beta_e(\theta') &= 1 -\frac{1}{N}\sum_{j=0}^{N-1}\langle\beta_j(\theta')|\hat{\Pi}_j^{\rm\textsc{me}}|\beta_j(\theta')\rangle
\nonumber \\
 &= 1-\frac{1}{N}(1+\sin 2\theta').
\label{eq:p_e_beta}
\end{align}

\subsection{Optimal FRIO discrimination of symmetric states as a two-step process}
\label{subsec:FRIO}

The optimal FRIO discrimination of the $N$ symmetric states will be decomposed into the two steps outlined above: an optimal state separation followed by a ME measurement on the successfully separated inputs. To see this, we represent this two-step process as an $(N+1)$-outcome POVM $\{\hat{\Pi}_0,\ldots,\hat{\Pi}_{N-1},\hat{\Pi}_?\}$, with the corresponding detection operators given by $\hat{A}_j=(\hat{\Pi}_j^{\rm\textsc{me}})^{1/2}\langle\mathsf{v}|\hat{U}(\theta')|\mathsf{v}\rangle$,
$\hat{A}_?=\langle\mathsf{h}|\hat{U}(\theta')|\mathsf{v}\rangle$, for $j=0,\ldots,N-1$, so that $\hat{\Pi}_j=\hat{A}_j^\dag\hat{A}_j$ and $\hat{\Pi}_?=\hat{A}_?^\dag\hat{A}_?$. It is easy to verify that all the elements are positive semidefinite and satisfy $\sum_j\hat{\Pi}_j+\hat{\Pi}_?=\hat{I}$. Using this POVM and Eqs.~(\ref{eq:symmetric_beta}), (\ref{eq:Unitary}) and (\ref{eq:POVM_ME}), we obtain the average probabilities of Eqs.~(\ref{eq:PePcQ}) as functions of $\theta'$, the polar angle after separation: 
\begin{subequations}     \label{eq:PePcQ-2}
\begin{align}
P_e(\theta')&=p_s(\theta')p_e^\beta(\theta'),\\
P_c(\theta')&=p_s(\theta')\left[1-p_e^\beta(\theta')\right], \\
Q(\theta')&=p_f(\theta'),  \label{eq:Q_FRIO}
\end{align}
\end{subequations}
where $p_{s,f}(\theta')$ and $p_e^\beta(\theta')$ are given by Eqs.~(\ref{eq:p_succ}) and (\ref{eq:p_e_beta}) respectively. Equations~(\ref{eq:PePcQ-2}) make explicit the connection between the two steps and the optimal FRIO discrimination. As we saw earlier, when the separation fails every input state is transformed as $|\alpha_j(\theta)\rangle\rightarrow|0\rangle$. In this case, an attempt to discriminate the inputs will lead to an inconclusive answer and, thereby, the rate of inconclusive results equals the failure probability in the separation. On the other hand, a successful separation will lead to a conclusive outcome for the discrimination attempt, and indeed we see that $P_e(\theta')+P_c(\theta')=p_s(\theta')$. The fixed value of $Q$ in the range $0\leqslant Q\leqslant Q^{\rm\textsc{mc}}$ is settled in the state separation stage by setting the angle $\theta'$ in the range $\theta\leqslant\theta'\leqslant\pi/4$. From Eqs.~(\ref{eq:p_succ}) and (\ref{eq:Q_FRIO}) we have $Q(\theta)=0$ and
\begin{equation}    \label{eq:Q_MC}
Q^{\rm\textsc{mc}}=Q(\pi/4)=\cos 2\theta,
\end{equation}
which is the minimum rate of inconclusive results for the optimal MC (or UD) discrimination of $N$ equally likely symmetric states of a qubit \cite{Chefles98-2,Croke06,Jimenez11}.

Finally, using Eqs.~(\ref{eq:p_succ}), (\ref{eq:p_e_beta}), (\ref{eq:PePcQ-2}), (\ref{eq:Q_MC}), and doing some algebra, we obtain the minimum error rate as a function of the fixed rate of inconclusive outcomes:
\begin{equation}     \label{eq:P_Q_FRIO}
P^{\rm min}_e(Q)=\frac{1}{N}\left[(N-1)\bar{Q} -\sqrt{\bar{Q}^2-(Q-Q^{\rm\textsc{mc}})^2}\right],
\end{equation}
where $\bar{Q}=1-Q$. This expression is in agreement with previous results in the literature \cite{Chefles98,Bagan12,Herzog12} and we can use it to check that the ME and optimal MC strategies emerge as particular cases of the optimal FRIO for $Q=0$ and $Q=Q^{\rm\textsc{mc}}$, respectively. In the first case, we have $P_e^{\rm min}(0)=1-\frac{1}{N}(1+\sin 2\theta)$, which is the ME bound for the input symmetric states $\{|\alpha_j(\theta)\rangle\}$ [see Eq.~(\ref{eq:p_e_beta}) for $\theta'=\theta$]. In the second case, $P_e^{\rm min}(Q^{\rm\textsc{mc}})=(1-Q^{\rm\textsc{mc}})(1-C)$, where $C=2/N$ is the maximum confidence achieved for each of the $N$ symmetric states \cite{Jimenez11}; for $N=2$ (optimal UD strategy), we have $P_e^{\rm min}(Q^{\rm\textsc{ud}})=0$. It is interesting to see that when the number of states to be discriminated increases, the chance of getting correct results decreases; in the limit $\lim_{N\rightarrow\infty}P_e^{\rm min}(Q)=1-Q$, so only erroneous or inconclusive answers can be obtained.

\section{Experiment}    \label{sec:Experiment} 

\begin{figure}[t]
\centerline{\includegraphics[width=1\columnwidth]{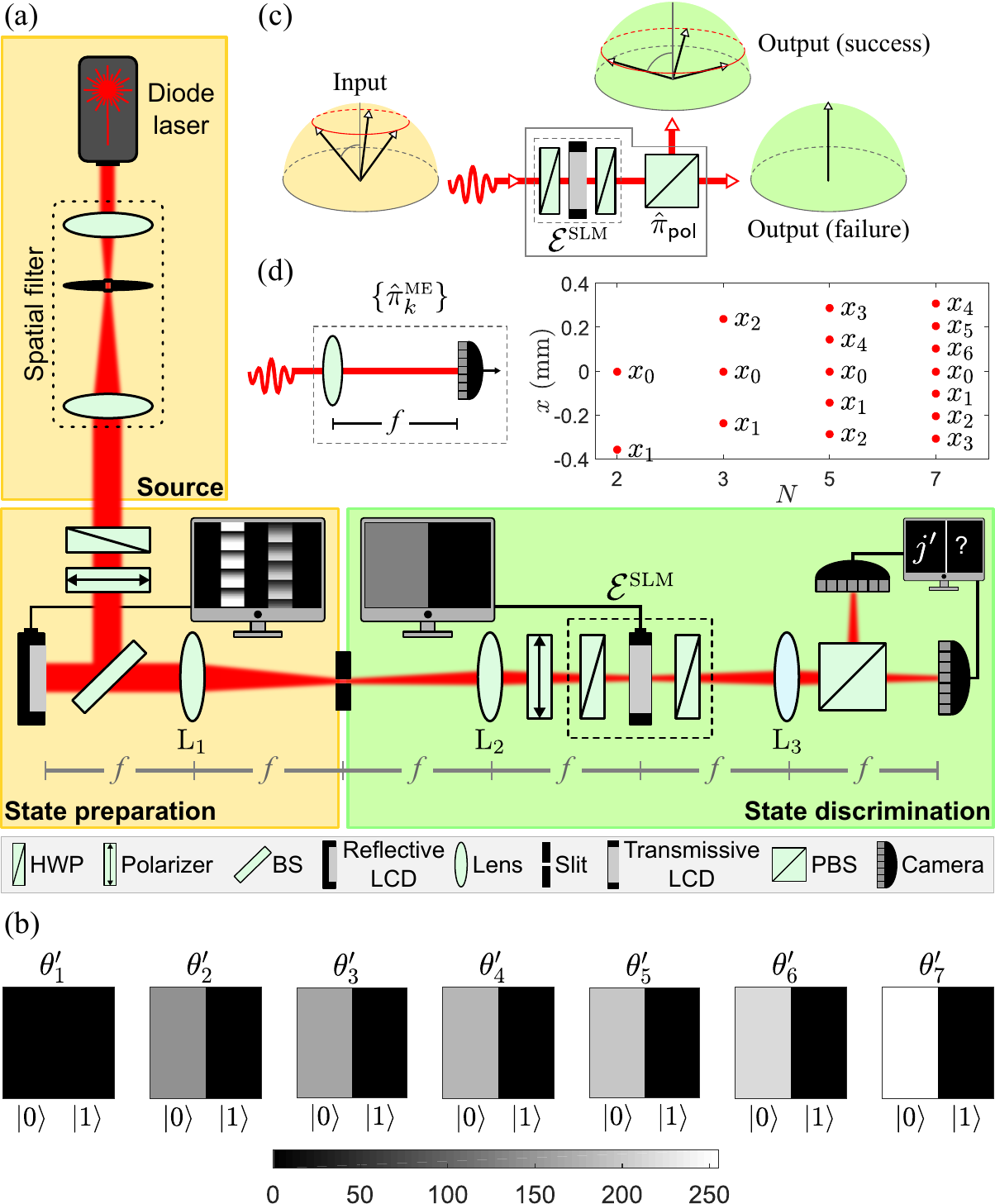}}
\caption{\label{fig:setup} (a) Experimental setup (see text for details). HWP: half-wave plate; BS: beam splitter; LCD: Liquid crystal display; PBS: polarizing beam splitter; L$_1$--L$_3$: spherical lenses with focal length $f=30\,$cm. The dashed box highlights the spatial light modulator (SLM) that performs the operation $\mathcal{E}^{\rm\textsc{slm}}$ described in the text. (b) Computer-generated masks addressed to the transmissive LCD to assist the state separation $|\alpha_j(\theta)\rangle\rightarrow|\beta_j(\theta'_t)\rangle$, where $\theta=19.5^\circ$ and $\{\theta'_t\}_{t=1}^7$ are given in Table~\ref{tab:theta}. (c) Arrangement for state separation: the SLM couples the path modes with the polarization, which is then measured in the $\{|\mathsf{h}\rangle,|\mathsf{v}\rangle\}$ by the PBS; the separation succeeds (fails) with the projection onto $|\mathsf{v}\rangle$ ($|\mathsf{h}\rangle$). (d) ME measurement to discriminate $N$ symmetric states of path encoded qubits [see Eq.~(\ref{eq:ME_proj})]: an array of $N$ pointlike detectors, at the focal plane of a lens, is distributed along the transverse positions $\{x_k\}_{k=0}^{N-1}$ given by Eq.~(\ref{eq:x_detector}); the panel shows these positions in our experiment.  }
\end{figure}

The experimental setup to demonstrate the optimal FRIO measurement is illustrated in Fig.~\hyperref[fig:setup]{\ref{fig:setup}(a)}. Next, we describe each section of our optical implementation, namely, state preparation, separation and discrimination. It is important to stress that, like most optical tests of quantum state discrimination \cite{Barnett97,Clarke00,Prosser17,Huttner96,Clarke01,Mohseni04,Prosser21,Mosley06}, our implementation explores the isomorphism between the state of an optical field generated by a laser source and a quantum state, as explained below.\footnote{As discussed in previous works \cite{Prosser17,Prosser21}, our implementation could be made truly quantum only by replacing the laser source by a single photon source and the cameras by detector arrays with single photon counting capability.}

\subsection{Source and state preparation}

Our light source consisted of a 687~$\!$nm single-mode diode laser. The beam is initially sent through a spatial filter that cleans, expands and collimates its spatial profile, generating an approximate plane wave field. Then, it passes through a half-wave plate (HWP) followed by a polarizer: the former acts as a variable attenuator and the latter provides a clean vertical polarization for the field.

The incoming beam produced at the source is driven to a normal incidence at a reflective liquid crystal display (LCD, Holoeye PLUTO) working as a programmable phase-only spatial light modulator (SLM). This device is addressed with a computer-generated mask given by an array of two blazed diffraction gratings [a typical mask is shown at the computer screen in the state preparation stage of Fig.~\hyperref[fig:setup]{\ref{fig:setup}(a)}]. For a display with pixels $8\, \mu$m wide, the gratings have period, width and center-to-center separation of 12, 18, and 36 pixels, respectively. The SLM imprints the mask information into the phase profile of the beam; the modulated beam is then transmitted through the spherical lens L$_1$, and a slit diaphragm at its focal plane filters the first diffraction order. The filtered beam is given by a coherent superposition of two non-overlapping path modes generated by the gratings. These modes, represented by $|0\rangle$ and $|1\rangle$, are modulated by complex coefficients with magnitude and phase defined by the phase depth and lateral displacement of the gratings, respectively (see Refs.~\cite{Prosser13,Varga14}). The state of the field emerging from this process is equivalent to a stream of single photons prepared in a given pure state on a two-dimensional space spanned by the two path modes. In particular, we use this method to prepare the symmetric qubit states $|\alpha_j(\theta)\rangle$ given by Eq.~(\ref{eq:symmetric_alpha}).

\subsection{State separation}
\label{subsec:QSS_exp}

For the state separation  $|\alpha_j(\theta)\rangle\rightarrow|\beta_j(\theta')\rangle$ we use the light polarization as an ancilla qubit. As both path modes in the preparation stage are vertically polarized, the input state is given by $|\alpha_j(\theta)\rangle|\mathsf{v}\rangle$. Thus, to implement the controlled unitary given by Eq.~(\ref{eq:Unitary_coupling}), we must have the path-encoded qubit acting as the control for a transformation on the polarization one: if the path mode is $|0\rangle$, a phase $\varphi$ is added and the polarization is rotated as $|\mathsf{v}\rangle\rightarrow|\mathsf{p}(\theta')\rangle=\sqrt{1-\xi^2(\theta')}|\mathsf{h}\rangle+\xi(\theta')|\mathsf{v}\rangle$; otherwise, it is left unchanged. 

The transformation on the ancilla, denoted by $\mathcal{E}^{\rm\textsc{slm}}$, is performed by a programmable SLM composed of a transmissive LCD (Holoeye LC 2012) sandwiched between two HWPs with fixed orientations, as shown in the dashed box of Fig.~\hyperref[fig:setup]{\ref{fig:setup}(a)} (see Appendix~\ref{app:SLM} for a brief description of the SLM characterization). This device modulates the polarization and phase of the incoming light as a function of the gray level ($\mathsf{gl}=0,\ldots,255$) displayed onto each pixel of the LCD \cite{Moreno03}. Given a vertically polarized input beam, for $\mathsf{gl}=0$, it acts as an identity operation. On the other hand, for $\mathsf{gl}>0$, its action imprints a phase shift and approaches the desired polarization rotation.\footnote{Unwanted effects present in the SLM, such as depolarization, prevents it from working exactly as the required unitary polarization rotation. A full characterization of these effects demands a process tomography of the device, which is beyond the scope of the present work. We shall present this study elsewhere.} Now, to make this a controlled operation, each path mode is imaged onto one of the halves of the LCD screen by a 4$f$ optical system formed by the lenses L$_1$ and L$_2$, as shown in Fig.~\hyperref[fig:setup]{\ref{fig:setup}(a)} (the polarizer before the SLM is used only to ensure a pure vertical polarization for each mode). The mode $|0\rangle$ ($|1\rangle$) goes through the left (right) half which is addressed with a homogeneous computer-generated mask with $\mathsf{gl}>0$ ($\mathsf{gl}=0$) [a typical mask is shown at the corresponding computer screen of Fig.~\hyperref[fig:setup]{\ref{fig:setup}(a)}]. In this way, the SLM will only act in the path mode $|0\rangle$. The relationship between the gray level at the LCD with the target separation angles and the phase shifts for the input symmetric states are discussed in the Appendix~\ref{app:thetaphi}. The values of these parameters used in our experiment are specified in Table~\ref{tab:theta}; the corresponding masks to implement the intended transformations are shown in Fig.~\hyperref[fig:setup]{\ref{fig:setup}(b)}.

To conclude this stage, a polarization projection, $\hat{\pi}_{\rm\mathsf{pol}}$, is performed on the basis $\{|\mathsf{h}\rangle,|\mathsf{v}\rangle\}$ with a polarizing beam splitter (PBS). The vertically polarized component of the state is reflected by the PBS, resulting in a successful separation, while the horizontal component, associated with a failure, is transmitted by the PBS, as sketched in Fig.~\hyperref[fig:setup]{\ref{fig:setup}(c)}.

\begin{table}[t]
\caption{\label{tab:theta} Starting with a set of $N$ symmetric states characterized by $\theta=19.5^\circ$ (this choice is explained in the Appendix~\ref{app:thetaphi}), the target separation angles in our experiment are specified in th second column. The third column shows the required gray level at the left half of the LCD screen to implement the intended separation; the fourth column shows the phase shifts introduced by the SLM with the addressed gray level.}
\begin{tabular*}{\linewidth}{@{\extracolsep{\fill}} cccc}
\hline\hline\\[-2mm]
$t$ & Separation angle  & Gray level  & Phase shift  \\[1mm]
& $\{\theta'_t\}$ & $\{\mathsf{gl}_t\}$ & $\{\varphi_t\}$ \\[1mm]
\hline\\[-2mm]
1 & $19.5^\circ$ & 0 & 0 \\
2 & $22.6^\circ$ & 142 & $0.23\pi$ \\
3 & $25.5^\circ$ & 163 & $0.32\pi$ \\
4 & $29.5^\circ$ & 180 & $0.40\pi$ \\
5 & $34.2^\circ$ & 195 & $0.48\pi$ \\
6 & $40.0^\circ$ & 214 & $0.56\pi$ \\
7 & $45.0^\circ$ & 255 & $0.61\pi$  \\[1mm]
\hline\hline
\end{tabular*}
\end{table}

\subsection{Minimum-error measurement}

The separated states $\{|\beta_j(\theta')\rangle\}$ from the previous step must now be discriminated with a ME measurement given by the $N$-outcome POVM of Eq.~(\ref{eq:POVM_ME}). From Naimark theorem, this POVM can be extended to a projective measurement on an larger Hilbert space \cite{PeresBook}. To see this, let $\{|k\rangle\}_{k=0}^{N-1}$ be an orthonormal basis spanning an $N$-dimensional space $\mathcal{H}_N$. 
By applying the quantum Fourier transform  $\hat{\mathcal{F}}_N=\frac{1}{\sqrt{N}}\sum_{m,n=0}^{N-1}\omega^{mn}|m\rangle\langle n|$, we generate a conjugate orthonormal basis that can be written as
\begin{align}
|\mu_k\rangle &= \hat{\mathcal{F}}_N|k\rangle \nonumber \\
&= \sqrt{\frac{2}{N}}|u_k\rangle + \frac{1}{\sqrt{N}}\sum_{m=2}^{N-1}\omega^{mk}|m\rangle,
\end{align}
where $|u_k\rangle$ is given by Eq.~(\ref{eq:symmetric_uniform}). From Eqs.~(\ref{eq:symmetric_beta}) and (\ref{eq:POVM_ME}), it is straightforward to show that $\langle\beta_j(\theta')|\hat{\Pi}_k^{\rm\textsc{me}}|\beta_j(\theta')\rangle=|\langle\mu_k|\beta_j(\theta')\rangle|^2$. Therefore, the projective measurement
\begin{equation}     \label{eq:ME_proj}
\hat{\pi}_k^{\rm\textsc{me}}=|\mu_k\rangle\langle\mu_k|=\hat{\mathcal{F}}_N|k\rangle\langle k|\hat{\mathcal{F}}_N^{-1}
\end{equation}
in the larger space $\mathcal{H}_N$, implements, in the qubit space $\mathcal{H}$, the POVM (\ref{eq:POVM_ME}) for the required ME discrimination.

Here, this projective measurement is performed by an array of $N$ pointlike detectors at the focal plane of a spherical lens, as sketched in Fig.~\hyperref[fig:setup]{\ref{fig:setup}(d)}. The lens performs an optical Fourier transform, and the $k$th detector in the array, located at the transverse position $x_k$, postselects the state \cite{Prosser17}
\begin{equation}
|\mu(x_k)\rangle=\frac{1}{\sqrt{N}}\sum_{l=0}^{N-1}\omega^{x_kNl\Delta/\lambda f}|l\rangle,
\end{equation}
where $\Delta$ is the distance between the path modes, $\lambda$ is the light wavelength, and $f$ the lens focal length. Thus, the ME measurement is implemented by distributing the detectors along the transverse positions
\begin{equation}    \label{eq:x_detector}
x_k=-\frac{\lambda f m_k}{N\Delta} \;\; \Rightarrow \;\; |\mu(x_k)\rangle\langle\mu(x_k)|=\hat{\pi}_k^{\rm\textsc{me}},
\end{equation}
where $k=0,\ldots,N-1$ and $m_k=k$ if $k\leqslant N/2$ or $m_k=k-N$, otherwise. The panel in Fig.~\hyperref[fig:setup]{\ref{fig:setup}(d)} shows these positions for $N=2,3,5,7$ symmetric states, obtained with our experimental parameters $\lambda=687\,$nm, $f=30\,$cm and $\Delta=288\,\mu$m.

We use CMOS cameras (Thorlabs DCC1545M) at the focal plane of the lens L$_3$ at both outputs of the PBS, as shown in Fig.~\hyperref[fig:setup]{\ref{fig:setup}(a)}. From each camera we select $N$ pixels (for a pixel size of $5.2\,\mu$m) located at the positions shown in Fig.~\hyperref[fig:setup]{\ref{fig:setup}(d)}. With the detections at the reflected arm, we obtain the error rates in the discrimination of the successfully separated states; with the detection at both arms, we obtain the rate of inconclusive outcomes, as explained next.

\section{Experimental results}

To carry out the experiment, we first define the number of states to be discriminated, $N$, and the fixed rate of inconclusive results, $Q(\theta')$, which is settled by the target separation angle $\theta'_t$. For a given $N$ and $\theta'_t$, the input states, given by Eq.~(\ref{eq:symmetric_alpha}), are prepared with $\theta=19.5^\circ$ (see Appendix~\ref{app:thetaphi}) and a relative phase $\varphi=\varphi(\theta'_t)\equiv\varphi_t$ shown in Table~\ref{tab:theta}. The inputs are prepared one at a time, each one with its corresponding mask displayed at the reflective LCD. All of them are subjected to the same operation by the SLM, defined by a fixed mask addressed to the transmissive LCD, according to $\theta'_t$ [see Fig.~\hyperref[fig:setup]{\ref{fig:setup}(b)}]. Finally, the cameras at both outputs of the PBS (success and failure, $\ell=s,f$) record the intensity distributions, $I_{j}^\ell(x,y)$, for each input state $j$. For each distribution, we subtract the background noise and integrate over the transverse direction $y$, obtaining $I_{j}^\ell(x)$, which will be used to characterize the state separation and estimate the probabilities in the discrimination process, as described next. 

\subsection{Characterizing the separated states}
\label{subsec:QSS_charact}

\begin{figure*}[t]
\centerline{\includegraphics[width=1\textwidth]{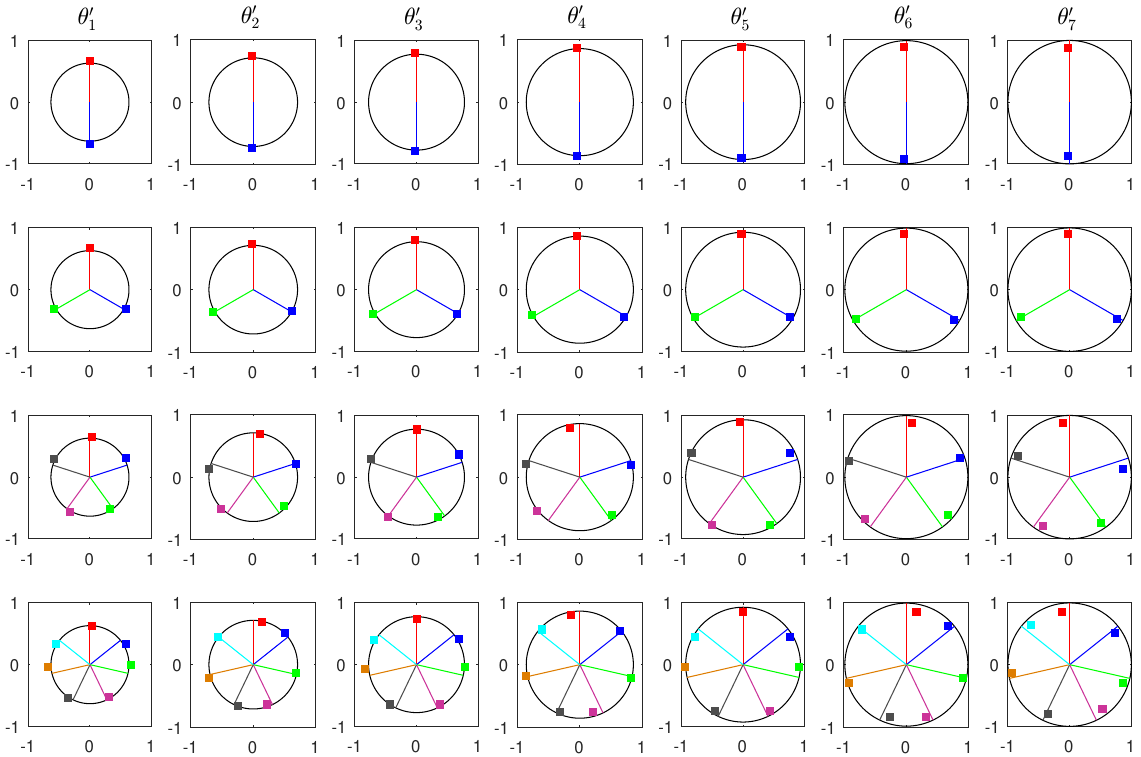}}
\caption{\label{fig:Bloch_exp} Characterization of the state separation process for $N=2,3,5$ and $7$ symmetric states (rows) with the target separation angles $\{\theta'_t\}_{t=1}^7$ (columns) given in Table~\ref{tab:theta}. The black circles represent the parallels of the Bloch sphere of radius $\sin 2\theta'_t$. The colored straight lines represent the target output symmetric states, where $|\beta_0(\theta'_t)\rangle$ is the red line (positive $y$-axis) and $\{|\beta_j(\theta'_t)\rangle\}_{j=1}^{N-1}$ are the remaining lines along the clockwise direction. The separated states, characterized experimentally, are shown as squares with the same colors of the target states.} 
\end{figure*}

First, we use the intensity distribution of a successfully separated state, $I_{j}^s(x)$, to characterize this state and the separation process itself. The measured $I_{j}^s(x)$ consists of an interference pattern between the two path modes. By applying a least square fitting to the data we obtain the visibility, $V_j$, and phase shift, $\phi_j$, of this pattern. With these parameters, we write the experimentally separated state for a given angle $\theta'_t$, as the following density matrix (see Appendix~\ref{app:QSS})
\begin{equation}    \label{eq:QSS_exp}
\hat{\rho}_j(\theta'_t)=
\frac{1}{2}\left[
\begin{array}{cc}
2\cos^2\theta'_t & e^{-i\phi_j}V_j \\[2mm]
 e^{i\phi_j}V_j & 2\sin^2\theta'_t
\end{array}
\right].
\end{equation}
In a perfect separation, the output state would be given by Eq.~(\ref{eq:symmetric_beta}), for which $\phi_j=2\pi j/N$ and $V_j=\sin 2\theta'_t$ $\forall j$. The target separation angle $\theta'_t$ sets the parallel of the Bloch sphere of radius $\sin 2\theta'_t$ [see Fig.~\hyperref[fig:Bloch]{\ref{fig:Bloch}(b)}]. Thus, by replacing $\theta'_t$ and the corresponding measured parameters $\phi_j$ and $V_j$ in Eq.~(\ref{eq:QSS_exp}), we obtain the location of $\hat{\rho}_j(\theta'_t)$ in the plane containing the parallel, which can be compared with the location of $|\beta_j(\theta'_t)\rangle$.

Figure~\ref{fig:Bloch_exp} shows the results obtained from this analysis for $N=2,3,5$ and $7$ symmetric states, and the separation angles $\{\theta'_t\}_{t=1}^7$ given in Table~\ref{tab:theta}, arranged in the rows and columns, respectively. The black circles represent the parallels of the Bloch sphere set by $\theta'_t$. The colored straight lines locate the target output symmetric states, where $|\beta_0(\theta'_t)\rangle$ is the red line (positive $y$-axis) and $\{|\beta_j(\theta'_t)\rangle\}_{j=1}^{N-1}$ are the remaining lines along the clockwise direction. The states $\hat{\rho}_j(\theta'_t)$ are shown as square markers with the same colors of the target states.

\begin{figure*}[t]
\centerline{\includegraphics[width=1\textwidth]{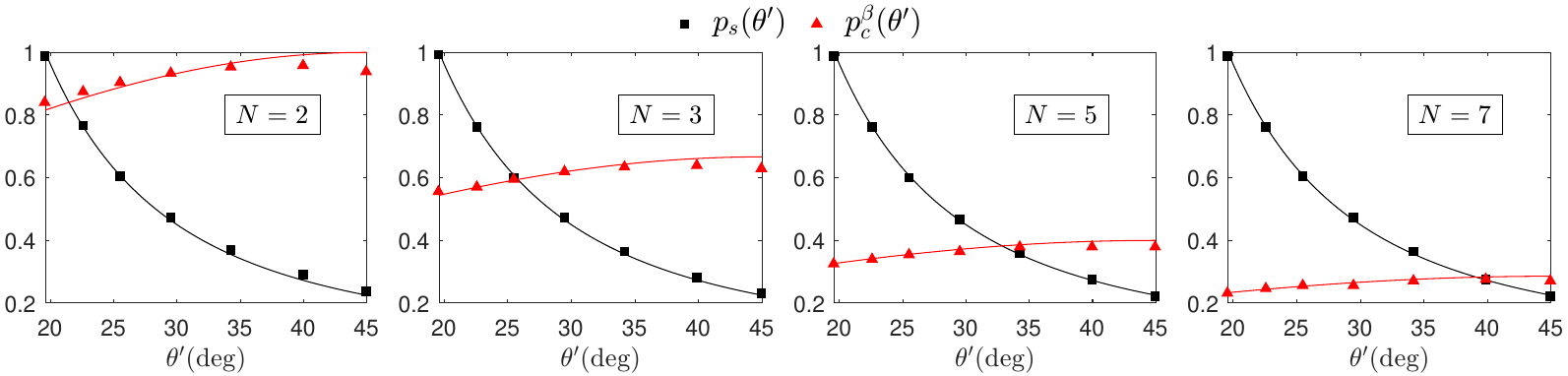}}
\caption{\label{fig:sep_me} Experimental results: average probabilities of successful state separation (black squares) and correct discrimination of the separated states (red triangles). The solid curves correspond to the optimal theoretical expectations $p_s(\theta')=1-p_f(\theta')$ and  $p_c^\beta(\theta')=1-p_e^\beta(\theta')$, given by Eqs.~(\ref{eq:p_succ}) and (\ref{eq:p_e_beta}), respectively. In both cases the standard deviations are of the order of $10^{-3}$. Thus, the error bars were smaller than the size of the data points and were not shown.}
\end{figure*}

In regard to the azimuthal angles $2\pi j/N$ that characterize the symmetric states, there is an excellent agreement between theory and experiment for $N=2$ and $3$. This agreement gets a little worse for $N=5$ and $7$, although it is still good. As $N$ increases, the adjacent states of a parallel become closer, which means that their relative phases are also closer and thus more susceptible to errors due to the finite phase resolution of the LCDs in both the preparation and separation stages. The radial location of $\hat{\rho}_j(\theta'_t)$ is determined by the measured visibility $V_j$. For a given $\theta'_t$, a visibility $V_j<\sin 2\theta'_t$ indicates loss of purity of the path-encoded states. In our experiment, this is observed more significantly for $\theta'_6$ and, specially, $\theta'_7$, as seen in Fig.~\ref{fig:Bloch_exp}. The main cause for this is the depolarization of the ancilla induced by the transmissive LCD (see Appendix~\ref{app:thetaphi}), which is more detrimental for the gray levels associated with these separation angles. As an effect, the error rate in the discrimination increases since the ME measurement relies on the interference of the path modes.\footnote{Note that the error probability in Eq.~(\ref{eq:p_e_beta}) can be written as $1-(1+V_j)/N$.} On the other hand, in a few cases we also obtained a slight deviation $V_j>\sin 2\theta'_t$ (e.g., $N=2$ and $\theta'_1$), which we attribute to inaccuracies in the preparation of the input states.\footnote{As $\theta'_1=\theta$, the SLM does not change the ancilla state for a zero gray level.} In this case, the consequence is the decreasing of the expected error rate. 

Despite experimental imperfections, the state separation increased the distinguishability of the input states in all instances $\theta=\theta'_1\rightarrow\theta'_t$ for $t=2,\ldots,7$. The gradual increasing of the distinguishability, accounted by the measured visibilities, was also observed from one step to another, i.e., $\theta'_t\rightarrow\theta'_{t+1}$, with the exception of $\theta'_6\rightarrow\theta'_7$ for which there is a slight decreasing in the visibilities for the reasons discussed above.

\subsection{The probabilities in FRIO discrimination}
\label{subsec:FRIO_probs}

The experimental success probability in the state separation, $[p_s]_{\rm expt}$, is obtained from the intensity distributions at both outputs of the PBS. First, we integrate them over $x$, $I_j^\ell=\sum_xI_j^\ell(x)$, and then, for each input $j$, we compute $p_{sj}=I_j^s/(I_j^s+I_j^f)$. Finally, we average this over all the inputs, obtaining $[p_s]_{\rm expt}=\sum_jp_{sj}/N$. Figure~\ref{fig:sep_me} shows the success probability as a function of the separation angles for each $N$. There is a good agreement between the experimental results (black squares) and  the optimal theoretical expectations (black curves) given by Eq.~(\ref{eq:p_succ}). 

The experimental probabilities of correctly identifying the separated states, $[p_c^\beta]_{\rm expt}$, are computed from the intensity distributions $I^s_j(x)$ as follows. First, the fits used to characterize state separation are also used to determine offsets to correct the $x$-axis, thus ensuring greater accuracy in locating the single-pixel detectors (see Appendix~\ref{app:QSS}). Then, for a given input state $j$ we collect the intensities at the $N$ pixels located in the positions given by Eq.~(\ref{eq:x_detector}) and apply a small compensation for the detection efficiency due to diffraction, yielding $\{\mathcal{I}^s_j(x_k)\}_{k=0}^{N-1}$.\footnote{The compensated intensity is given by $\mathcal{I}^s_j(x_k)=I^s_j(x_k)/\chi_k$, where the compensation factor $\chi_k={\rm sinc}^2(\pi\Lambda x_k/\lambda f)$ depends only on the detector position \cite{Prosser17}; here $\Lambda=144\,\mu$m is the width of the path mode. } From these intensities we obtain $p_{jk}=\mathcal{I}^s_j(x_k)/\sum_{l=0}^{N-1}\mathcal{I}^s_j(x_l)$, namely, the  conditional probabilities of correct ($j=k$) or erroneous ($j\neq k$) identifications of the separated states. Finally, the experimental average rate of correct discrimination will be given by $[p_c^\beta]_{\rm expt}=\sum_{j=0}^{N-1}p_{jj}/N$. These results are shown in Fig.~\ref{fig:sep_me} (red triangles) as a function of the separation angles for each $N$; the red curves correspond to the optimal theoretical predictions, $p_c^\beta(\theta')=1-p_e^\beta(\theta')$, given by Eq.~(\ref{eq:p_e_beta}). In general, there is a good agreement between theory and experiment; the observed discrepancies are mainly due to the issues pointed out in the state separation stage, as discussed in the previous subsection. Despite them, one clearly observes that the rate of correct results for the separated states increases with $\theta'$, reflecting the increase in their distinguishability.

\begin{figure}[t]
\centerline{\includegraphics[width=1\columnwidth]{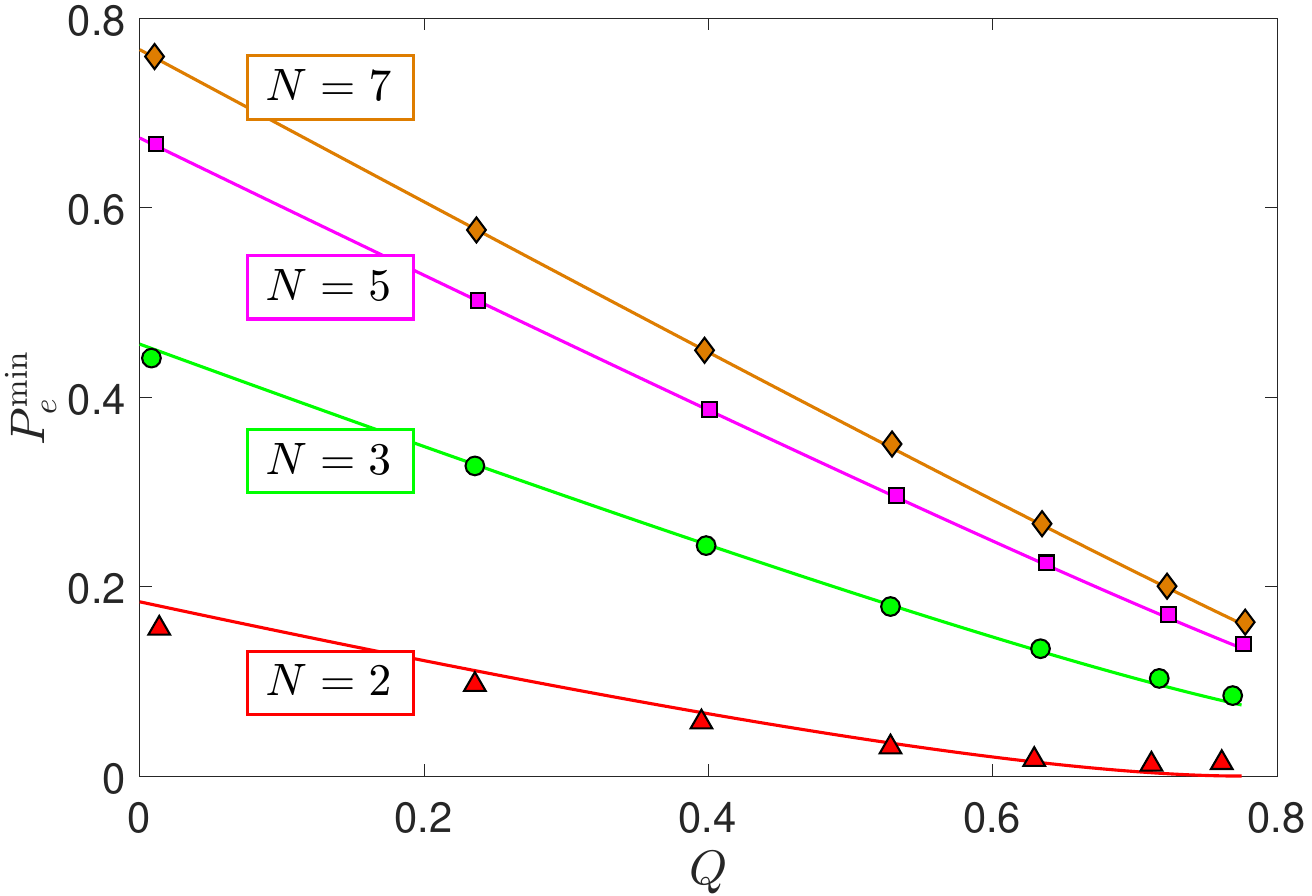}}
\caption{\label{fig:PeQ} Average error probability, $P_e^{\rm min}$, as a function of the fixed rate of inconclusive outcomes, $Q$. The markers represent the experimental results and the solid curves the optimal theoretical expectations given by Eq.~(\ref{eq:P_Q_FRIO}). The insets show the number of states, $N$. Again, the error bars were not shown for the same reasons described in Fig.~\ref{fig:sep_me}.}
\end{figure}

From these data, we can calculate the FRIO figure of merit using Eqs.~(\ref{eq:PePcQ-2}). The experimental error rates and fixed rates of inconclusive results will be given, respectively, by $[P_e]_{\rm expt}=[p_s]_{\rm expt}(1-[p_c^\beta]_{\rm expt})$ and $Q_{\rm expt}=1-[p_s]_{\rm expt}$. In Fig.~\ref{fig:PeQ} we plot the former as a function of the latter (markers) for each $N$ indicated in the insets. The curves of same color as the markers correspond to the optimal theoretical prediction, $P_e^{\rm min}(Q)$, given by Eq.~(\ref{eq:P_Q_FRIO}), and show good agreement with the data. It can be seen that this agreement becomes better as the number of input states increases. As discussed in Sec.~\ref{subsec:FRIO}, this occurs because as $N$ increases the errors and inconclusive results are predominant in the discrimination process. In this way, experimental imperfections independent of $N$ become less noticeable. Still, we can see from Fig.~\ref{fig:PeQ} that our FRIO measurement scheme for discriminating between $N$ equally likely symmetric states, closely reaches the minimum error rate for a fixed value of $Q$, and interpolates between the ME discrimination ($Q=0$) and optimal UD (for $N=2$) or MC (for $N>2$) strategies ($Q=0.7771$).

\section{Discussion and conclusion}

The decomposition of the optimal FRIO measurement as a two-step process gives a clearer and more instructive view on how the probabilistic discrimination strategies work in practice. In the first step, a quantum map is applied to increase the distinguishability between the input states with a minimum probability of failure. The increase in distinguishability depends on the desired level of confidence in the discrimination that follows, and by fixing this increase, we are fixing the rate of inconclusive results. Then, in the second step, the successfully transformed states are discriminated with a minimum-error measurement. When the increase in distinguishability is zero, we have a ME measurement and when it is maximal we have an optimal UD (for linearly independent states) or an optimal MC (for linearly dependent states) strategy, showing in a simple way how the optimal FRIO interpolates between those extreme strategies.

Here, we adopted this two-step approach and demonstrated, both theoretically and experimentally, the optimal FRIO discrimination between $N$ equally symmetric pure states of a qubit. Although our experiment has been carried out with a classical laser source, the results would not differ were it be done with a true single photon source and a detector array with single photon counting capability, as extensively discussed in previous works \cite{Prosser17,Prosser21}.

Our implementation employed two path modes of light to encode the symmetric states; the light polarization was used as an ancilla system to perform state separation, which controllably increased the distinguishability between the symmetric states. This transformation was implemented through a programmable spatial modulator, which gave us a fine control for the transition between ME measurement and optimal UD (or MC) measurements. In addition to show the optimal FRIO measurement for $N$ states of a qubit, rather than only two, our setup has an advantage over a previous FRIO implementation \cite{Gomez22} that used polarization-encoded qubits: it can be straightforwardly extended to high-dimensional qudits. We have all the ingredients for that: the qudit states can be encoded in $d$ path modes produced by an array of $d$ blazed diffraction gratings at the SLM \cite{Prosser13} [similarly to the two-dimensional case shown in the computer screen of Fig.~\href{fig:setup}{\ref{fig:setup}(a)}]. In addition, both extreme strategies, ME and UD,  have been demonstrated for discriminating path-encoded symmetric states of qudits \cite{Prosser17,Prosser21}. The FRIO measurement in this case requires the parametric state separation in the first step, a protocol introduced in Ref.~\cite{Prosser16}. This protocol also uses a two-dimensional ancilla and can be performed in a similar way to what we did here, using an SLM (the optimal UD \cite{Prosser21}, in particular, employed this transformation to implement the maximum separation). After that, we just implement the ME measurement on the successfully separated states following the method shown in \cite{Prosser17}. Therefore, our encoding enables a single setup where one can implement the most fundamental state discrimination strategies in dimensions much larger than two. The switching between these strategies is controlled just by tuning the transformation carried out by the SLM, which means to change the computer-generated mask addressed to a programmable LCD.

In conclusion, we have demonstrated an experimental implementation of the optimal FRIO measurement for discriminating between $N=2,3,5,$ and $7$ nonorthogonal states of a qubit. Our results clearly showed the gradual decreasing of the error rate with the increasing of the fixed rate of inconclusive outcomes, encompassing the extreme cases of ME and UD (or MC) measurements.

\begin{acknowledgments}
This work was supported by FAPEMIG, CNPq, and CNPq INCT-IQ (465469/2014-0). L. F. M. acknowledges financial support from CAPES - Finance Code 001. O. J. was supported by an internal grant from Universidad Mayor PEP I-2019020. A. D. was supported by FONDECYT Grants 1231940 and 1230586 and ANID  -- Millennium Science Initiative Program -- ICN17$_-$012.
\end{acknowledgments}

\appendix

\section{Calibration and configuration of the SLM for state separation}
\label{app:SLM}

In order to characterize the optical modulation properties of the transmissive LCD used in the state separation stage, we resort to the standard approach developed by Moreno \textit{et al.} \cite{Moreno03}. First, to characterize the polarization modulation, the LCD is sandwiched by a polarization state generator (PG) and polarization analyzer (PA), the former composed of a linear polarizer followed by a quarter-wave plate (QWP) and the later of a QWP followed by a linear polarizer. With this arrangement, we measure the light intensities as a function of the gray level addressed to the display by preparing and analyzing the polarization in the bases $\{\mathsf{h,v}\}$, $\{\pm 45^\circ\}$ and $\{\mathsf{R,L}\}$, which generates 36 measurements, each one for $\mathsf{gl}=0,\ldots,255$. Second, to characterize the phase modulation, we remove both PG and PA and illuminate each half of the LCD with a small vertically polarized beam generated by a double pinhole. This double beam is made to interfere at the focal plane of a spherical lens at the vertical output of a PBS, where the intensity pattern is recorded by a CMOS camera. By keeping the gray level at the left half of the LCD equal zero while changing it at the right half, we record 256 interference patterns. 

The SLM is built by sandwiching the LCD between a HWP$+$QWP (before) and QWP$+$HWP (after) with fixed orientations. With the dataset from the LCD characterization outlined above we can predict the configurations of the wave plates to obtain the desired light modulation. In our case, the imposed constraints were twofold: first, a vertically polarized light passing through the SLM must not suffer any net rotation for $\mathsf{gl}=0$ at the LCD and its intensity must decrease monotonically as $\mathsf{gl}$ increases; second, the phase shift introduced by the SLM as a function of $\mathsf{gl}$ must be equal at both outputs of a PBS. We made a numerical search for the configuration of the wave plates satisfying these constraints and found that only the HWPs were required before and after the LCD [see Fig.~\href{fig:setup}{\ref{fig:setup}(a)}] oriented at $28^\circ$ and $29^\circ$ from the vertical axis, respectively. The resulting modulation properties of this configuration are described in Appendix~\ref{app:thetaphi}.

\section{Target separation angle and phase shift as a function of the gray level at the LCD}
\label{app:thetaphi}

\begin{figure}[tbh]
\centerline{\includegraphics[width=1\columnwidth]{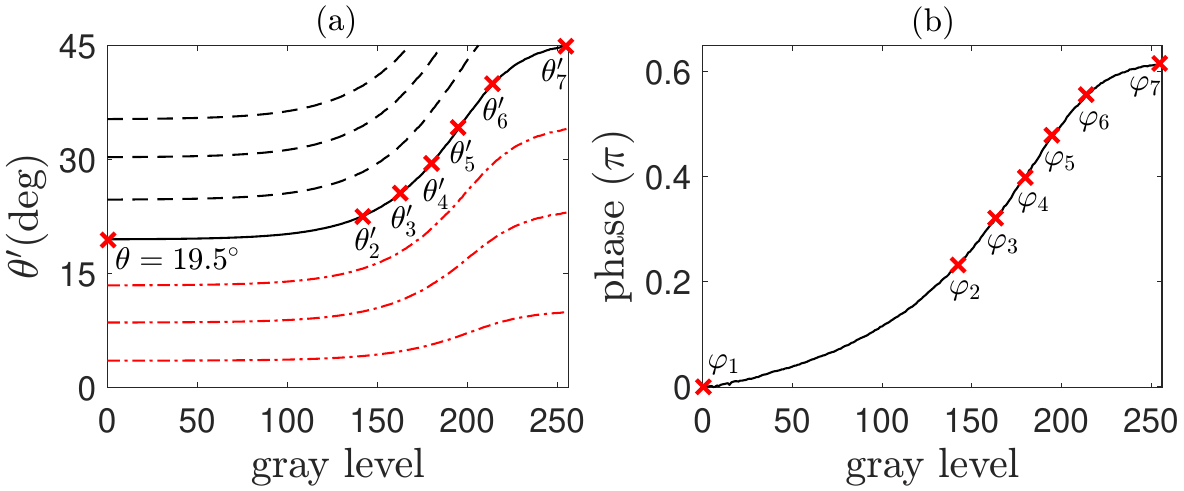}}
\caption{\label{fig:thetaphi} (a) Target separation angle and (b) phase shift as a function of the gray level addressed to the LCD. In (a), each curve corresponds to a given input angle $\theta=\theta'(0)$, which sets the range of accessible $\theta'$'s (see text for details). The red markers ``$\times$'' indicate the values of $\theta'_t$ and $\varphi_t$ used in our experiment (they are also specified in Table~\ref{tab:theta}). }
\end{figure}

In Sec.~\ref{subsec:QSS_exp} it was shown that to perform the optimal state separation on the path-encoded symmetric states, the programmable SLM must, for a given gray level at the LCD, adds a phase shift $\varphi$ in the mode $|0\rangle$ and rotates its polarization according to 
\begin{equation}    \label{eq:rotation_app}
|\mathsf{v}\rangle\rightarrow|\mathsf{p}(\theta')\rangle=\sqrt{1-\xi^2(\theta')}|\mathsf{h}\rangle+\xi(\theta')|\mathsf{v}\rangle. 
\end{equation}
The parameter $\xi(\theta')=\tan\theta\cot\theta'$ [see Eq.~(\ref{eq:Unitary_coupling})] sets the relationship between the input and output separation angles ($\theta$ and $\theta'$, respectively) with the required rotation of the ancilla to achieve that separation. Based on this, we can obtain the target output separation angles as a function of the gray level, namely $\theta'(\mathsf{gl})$. From Eq.~(\ref{eq:rotation_app}) we have $\xi^2(\theta')=|\langle\mathsf{v}|\mathsf{p}(\theta')\rangle|^2\equiv P_{\mathsf{v}}$, which is the probability of projecting the rotated ancilla onto the vertical polarization. The SLM is configured to implement this transformation as close as possible: unwanted effects such as depolarization \cite{Marquez08} prevents it from working exactly as we wish. In a simple model, this means that an input vertically polarized beam is actually transformed as 
\begin{equation}
|\mathsf{v}\rangle\rightarrow\hat{\rho}(\mathsf{gl})=
\bm{(}1-\epsilon(\mathsf{gl})\bm{)}|\mathsf{p}(\theta')\rangle\langle\mathsf{p}(\theta')|
+\epsilon(\mathsf{gl})\frac{\hat{I}}{2},
\end{equation}
with a desirable $\epsilon(\mathsf{gl})\rightarrow 0$ for $\mathsf{gl}=0,\ldots,255$. With this configuration we can extract $P_{\mathsf{v}}(\mathsf{gl})=\langle\mathsf{v}|\hat{\rho}(\mathsf{gl})|\mathsf{v}\rangle$ (see the appendix of Ref.~\cite{Prosser21}) and, assuming that the SLM implements the exact rotation, we obtain
\begin{equation}
\theta'(\mathsf{gl})=\arctan\left[\frac{\tan\theta}{\sqrt{P_{\mathsf{v}}(\mathsf{gl})}}\right].
\end{equation} 
In Fig.~\hyperref[fig:thetaphi]{\ref{fig:thetaphi}(a)}, we plot $\theta'(\mathsf{gl})$ for different values of the input angle, $\theta=\theta'(0)$. For $\theta<19.5^\circ$ (red dash-dotted curves), the accessible $\theta'$'s get below $45^\circ$, which makes it impossible to achieve maximum state separation. In this case, although FRIO could be implemented, we would not reach the MC measurement. On the other hand, for $\theta>19.5^\circ$ (black dashed curves), although $\theta'$ reaches $45^\circ$, the accessible angles vary within decreasing ranges. The best scenario to overcome these issues is provided by the input $\theta=19.5^\circ$, shown in the black solid curve, where the red markers ``$\times$''  indicate the target separation angles $\{\theta'_t\}_{t=1}^7$ (note that $\theta_1'=\theta$). For this reason, we have implemented the optimal FRIO measurement starting with symmetric states with $\theta=19.5^\circ$. Figure~\hyperref[fig:thetaphi]{\ref{fig:thetaphi}(b)} shows the phase shift imprinted by the SLM as a function of the gray level at the LCD, obtained through interferometric measurements \cite{Moreno03}. The red markers ``$\times$'' indicate the phases $\{\varphi_t\}_{t=1}^7$ corresponding to the gray levels used to achieve the target separation angles. The values of $\theta'_t$ and $\varphi_t$ used in our experiment are also given in Table~\ref{tab:theta}. 

\section{Characterizing the state separation of path-encoded qubits by interference measurements}
\label{app:QSS}

For a path-encoded qubit, the probability amplitude of a path mode $|m\rangle$ ($m=0,1$) at the focal plane of a lens is written, in position space, as \cite{Prosser11}
\begin{equation}
\langle x|m\rangle\propto \exp\bm{(}-i\kappa(1-2m)x\Delta\bm{)}\,{\rm sinc}(\kappa\Lambda x),
\end{equation}
where $\kappa=\pi/\lambda f$, with $\lambda$ denoting the light wavelength and $f$ the lens focal length; $\Delta$ is the distance between the modes and $\Lambda$ is their width. (The values of these parameters in our experiment were provided in the main text.) Hence, for a qubit in the symmetric state $|\beta_j(\theta')\rangle$ defined in Eq.~(\ref{eq:symmetric_beta}), the detection probability density at the focal plane will be given by
\begin{align}
I_j(x)&=\left|\langle x|\beta_j(\theta')\rangle\right|^2 \nonumber\\
&\propto {\rm sinc}^2(\kappa\Lambda x)\left[1+V(\theta')\cos(2\kappa x\Delta+\arg\omega^j)\right],
\end{align}
where $V(\theta')=\sin 2\theta'$. This expression represents an interference pattern modulated by the envelope ${\rm sinc^2}(\bullet)$; $V(\theta')\in[\sin 2\theta,1]$ is the visibility of this pattern whose fringes are displaced by $\arg\omega^j=2\pi j/N$. From this result, the density matrix of the symmetric state can be written as
\begin{equation}
|\beta_j(\theta')\rangle\langle\beta_j(\theta')|=
\frac{1}{2}\left[
\begin{array}{cc}
\displaystyle 2\cos^2\theta' & \omega^{-j}V(\theta') \\[2mm]
\displaystyle\omega^{j}V(\theta') & 2\sin^2\theta'
\end{array}
\right].
\end{equation}
Therefore, it can be seen that by measuring the interference pattern $I_j(x)$ and determining its visibility and phase shift, we obtain the off-diagonal terms of the density matrix; the terms in the diagonal are obtained by measuring the path-mode amplitudes.

The separation angle, $\theta'$, sets the parallel of radius $V(\theta')$ on the Bloch sphere, and the azimuthal angle $\arg\omega^j$ locates $|\beta_j(\theta')\rangle$ on this parallel, as seen in Fig.~\hyperref[fig:Bloch]{\ref{fig:Bloch}(b)}. For a given $\theta'$ and an azimuthal angle $\phi_j$, a path-encoded qubit state $\hat{\rho}_j(\theta')$ inside the parallel is mixed and its radial location is defined by the visibility of its interference pattern, $V_j\in[0,\sin 2\theta']$. Thus, this state can be written as
\begin{equation}     \label{eq:rho_appendix}
\hat{\rho}_j(\theta')=
\frac{1}{2}\left[
\begin{array}{cc}
\displaystyle 2\cos^2\theta' & e^{-i\phi_j}V_j \\[2mm]
\displaystyle e^{i\phi_j}V_j & 2\sin^2\theta'
\end{array}
\right],
\end{equation}
in accordance with Eq.~(\ref{eq:QSS_exp}).

The state separation process is characterized by measuring the intensity distribution of each successfully separated state, $I_j^s(x)$. For a given $N$ and a target separation angle $\theta'_t$, we apply a least square fitting to each measured $I_j^s(x)$ using the function
\begin{equation}
F_j(x)= I_j^{\rm max}{\rm sinc}^2(\kappa\Lambda x)\left[1+V_j\cos(2\kappa x\Delta+\phi'_j)\right],
\end{equation}
where $I_j^{\rm max}$ is a global proportionality constant for each distribution, $V_j$ is the visibility and $\phi'_j$ a phase shift. These are the parameters we obtain from the fitting. The visibility gives us the magnitude of the off-diagonal terms in Eq.~(\ref{eq:rho_appendix}). From the parameters $\{\phi'_j\}_{j=0}^{N-1}$ we compute a correction term $\phi_{\rm corr}=\pi(N-1)/N-\sum_{l=0}^{N-1}\phi'_l/N$ used to determine offsets to correct the $x$-axis as $x_{\rm corr}=x-\phi_{\rm corr}/2\kappa\Delta$. As described in Sec.~\ref{subsec:FRIO_probs}, this ensures greater accuracy in locating the single-pixel detectors which will implement the ME measurement. In addition, we also obtain the phases of the off-diagonal terms in Eq.~(\ref{eq:rho_appendix}) as $\phi_j=\phi'_j+\phi_{\rm corr}$, concluding the characterization of the separated states.

%

\end{document}